\documentclass[twocolumn, aps, prx, floatfix, superscriptaddress]{revtex4-2}

\usepackage{amsmath, amssymb, amsfonts}
\usepackage{graphicx}
\usepackage{hyperref}
\usepackage{xcolor}
\usepackage{braket}
\usepackage{physics}
\usepackage{booktabs}
\usepackage{siunitx}

\usepackage{bm}

\usepackage[font=small,labelfont=bf,format=plain]{caption}
\usepackage{subcaption}
\makeatletter
\long\def\@makecaption#1#2{%
  \par\vskip\abovecaptionskip
  \begingroup
    \small\rmfamily
    \leftskip\z@ \rightskip\z@ \parfillskip\@flushglue
    \parindent\z@ \textbf{#1:} #2\par
  \endgroup
  \vskip\belowcaptionskip}
\long\def\@caption#1[#2]#3{%
  \par
  \addcontentsline{\csname ext@#1\endcsname}{#1}%
    {\protect\numberline{\csname the#1\endcsname}{\ignorespaces #2}}%
  \begingroup
    \@parboxrestore
    \normalsize
    \@makecaption{\csname fnum@#1\endcsname}{\ignorespaces #3}\par
  \endgroup}
\makeatother

\newcommand{\QGT}{\mathbf{S}}

\begin{document}

\title{Neural Quantum States for Nuclear Magnetic Resonance Spectroscopy}

\author{Bharadwaj Chowdary Mummaneni}
\email{bharadwaj.chowdary.mummaneni@iao.fraunhofer.de}
\affiliation{Fraunhofer IAO, Nobelstraße 12, 70569 Stuttgart, Germany}

\author{Bo Xing}
\affiliation{Research Laboratory of Electronics, Massachusetts Institute of Technology, Cambridge, Massachusetts 02139, USA}
\affiliation{Institute of Advanced Intelligence and Computing (IAIC), Agency for Science, Technology and Research (A*STAR), Singapore 138632, Republic of Singapore}

\date{\today}

\begin{abstract}
Predicting a nuclear magnetic resonance (NMR) spectrum from first principles requires propagating a quantum state of dimension $2^N$ for $N$ coupled spins, which becomes intractable beyond larger $N$. We benchmark Neural Quantum States (NQS), a class of variational quantum states expressed as an artificial neural network, as an alternative representation for this problem. Using two propagation methods, the Time-Dependent Variational Principle (TDVP) and projected time-dependent Variational Monte Carlo (p-tVMC), we compute the $^1$H spectra of four ($2 \to 5$ spins) experimentally parameterized molecules. TDVP reproduces all line positions and intensities with average spectral mean squared errors of $<10^{-3}$; p-tVMC reproduces the same features, with accuracy determined by its per-step optimization parameters. One dominant obstacle to larger systems is the steep growth of the number of integration steps with spectral bandwidth, which can be removed by propagating in the interaction frame of the chemical-shifted Hamiltonian. Retaining the same accuracy, this reduces the number of integration steps roughly eightfold for the 3-spin system and by at least an order of magnitude for the 4- and 5-spin systems, and it enables a 14-spin molecule (sucrose) to be accurately propagated via Monte Carlo sampling.
\end{abstract}

\maketitle

\section{Introduction}

When a chemist synthesizes a new compound or isolates a natural product, one of the first questions is whether the molecule in hand is actually the one intended. Nuclear Magnetic Resonance (NMR) spectroscopy is the standard tool for answering this question. Each chemically distinct nucleus resonates at a characteristic frequency, its \emph{chemical shift}, and nuclei connected through chemical bonds interact via \emph{scalar couplings} that split resonances into multiplet patterns. Together, these features encode the connectivity, stereochemistry, and conformation of the molecule~\cite{ernst1987principles, levitt2008spin}. A proposed structure is confirmed when its predicted NMR spectrum matches the experimental one.

The first half of this prediction is now routine. Quantum-chemical methods, particularly Density Functional Theory (DFT), can compute NMR parameters (chemical shifts and $J$-couplings) from first principles~\cite{helgaker1999ab, lodewyk2012computational, wolinski1990efficient}, and programs such as \textsc{Gaussian}~\cite{gaussian16} and \textsc{ORCA}~\cite{neese2020orca} can perform these calculations as a standard part of structure elucidation. However, to obtain the actual spectrum, one must propagate $N$ coupled spin-$\frac{1}{2}$ nuclei under the effective Hamiltonian and study the frequency domain of the time-domain signal: a quantum many-body time evolution in a Hilbert space of dimension $2^N$, intractable beyond roughly $N = 15$--$20$~\cite{savostyanov2014exact}. Molecules with extended proton networks (natural products, peptides, polycyclic aromatics) routinely exceed this limit. Restricted state-space methods, implemented in the \textsc{Spinach} library, push the boundary considerably by discarding high-order spin correlations that remain essentially unpopulated in liquid-state experiments~\cite{kuprov2007polynomially, hogben2011spinach}, but their accuracy relies on truncation assumptions that must be validated for each coupling topology. Beyond the reach of these techniques, DFT-computed parameters cannot be turned into a spectrum, and the structural assignment remains incomplete.

Several strategies have been explored to overcome this exponential scaling. Nuclear spins map directly onto qubits~\cite{jones2024quantum, lloyd1996universal}, but current quantum hardware does not yet provide the gate fidelities and coherence times that are useful for NMR simulation~\cite{preskill2018quantum, khedri2024impact, fratus2025quantum}. Tensor network methods compress the quantum state when the entanglement dynamics is not extensive~\cite{orus2014practical, white2004real, vidal2004efficient}, but their effectiveness for long-range and higher-dimensional Hamiltonians has not been systematically assessed. A third, largely unexplored alternative is Neural Quantum States (NQS): variational wavefunctions whose amplitudes are given by a neural network~\cite{carleo2017solving}. Initially proposed for finding the ground-state energy, NQS have since been extended to real-time dynamics through the time-dependent variational principle (TDVP)~\cite{schmitt2020quantum, yuan2019theory} and the projected time-dependent variational Monte Carlo (p-tVMC) method~\cite{sinibaldi2023unbiasing,ZhangPoletti2025}. NQS can represent certain classes of quantum states with only polynomially many parameters~\cite{gao2017efficient, hibatallah2020recurrent, pfau2020ab,ZhangPoletti2023}, and their flexibility suits the complex coupling topologies that arise in NMR.

NMR poses a stringent test for any variational dynamics method. Unlike the ground-state energy, the quantity of interest is a frequency-domain response derived from a time-domain signal over the full simulation window, so even small phase errors accumulated during propagation appear as shifted or artificially broadened lines. Whether NQS can maintain the required accuracy over the hundreds to thousands of integration steps needed to resolve individual peaks can only be settled by direct numerical experiment.

In this work, we apply both TDVP and p-tVMC to four experimentally characterized $^1$H spin systems (2--5 coupled protons) and benchmark the resulting spectra against exact diagonalization. Both methods reproduce the correct line positions and intensities, with spectral mean squared errors of $<10^{-3}$, a proof of principle that NQS dynamics can compare DFT-computed parameters to experimentally obtained spectra. The principal bottleneck turns out to be the steep growth of the integration steps with system size and not neural network expressivity. We show that an exact transformation to the interacting frame can leave the spectrum unchanged reduces the steps needed at fixed accuracy roughly eightfold for the 3-spin system and by at least an order of magnitude for the 4- and 5-spin systems. This improvement in computation efficiency allows a 14-spin molecule (sucrose) to be propagated by Monte Carlo sampling.

\begin{figure*}[!ht]
    \centering
    \begin{subfigure}[b]{0.48\textwidth}
        \centering
        \includegraphics[width=0.6\textwidth]{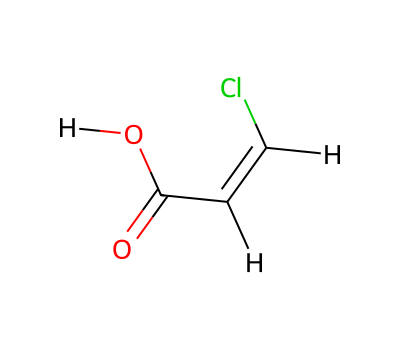}
        \caption{2-spin: \textit{cis}-3-Chloroacrylic acid}
    \end{subfigure}
    \hfill
    \begin{subfigure}[b]{0.48\textwidth}
        \centering
        \includegraphics[width=0.6\textwidth]{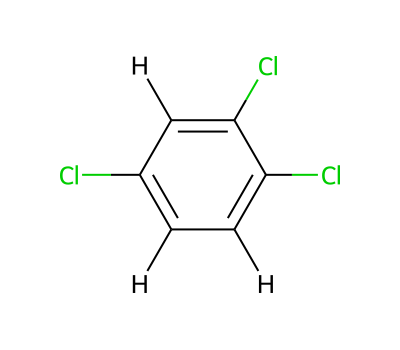}
        \caption{3-spin: 1,2,4-Trichlorobenzene}
    \end{subfigure}
    \\[1em]
    \begin{subfigure}[b]{0.48\textwidth}
        \centering
        \includegraphics[width=0.6\textwidth]{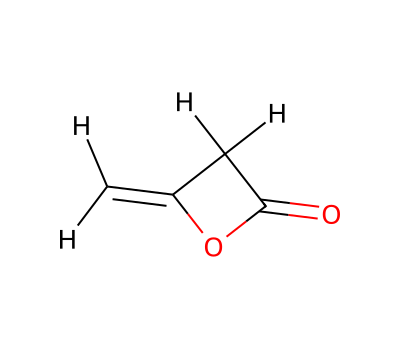}
        \caption{4-spin: 3-Methyleneoxetan-2-one}
    \end{subfigure}
    \hfill
    \begin{subfigure}[b]{0.48\textwidth}
        \centering
        \includegraphics[width=0.6\textwidth]{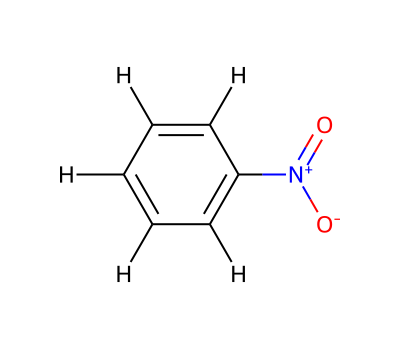}
        \caption{5-spin: Nitrobenzene}
    \end{subfigure}
    \caption{Molecular structures of the four spin systems studied in this work. Hydrogen atoms are shown explicitly; heteroatoms are colored (Cl: green, O: red, N: blue). From top-left to bottom-right: \textit{cis}-3-chloroacrylic acid (2 $^1$H), 1,2,4-trichlorobenzene (3 $^1$H), 3-methyleneoxetan-2-one (4 $^1$H), and nitrobenzene (5 $^1$H). NMR parameters are from experimental databases~\cite{dashti2017applications,hqs2023}.}
    \label{fig:molecules}
\end{figure*}

\section{Theoretical Model}

\subsection{From FID to Spectrum}

An NMR experiment records the \emph{free induction decay} (FID): after a short radiofrequency pulse tips the equilibrium spin magnetization into the transverse plane, the sample coil picks up the oscillating voltage induced by the precessing nuclear spins. Each chemically distinct proton precesses at a slightly different frequency, its \emph{chemical shift} $\delta_i$ (in ppm), set by how strongly the surrounding electron cloud shields it from the applied field. Where two protons are connected by a few bonds, their magnetic moments interact through bonding electrons via the \emph{scalar coupling} $J_{ij}$ (in Hz), which splits each resonance into a multiplet. Fourier-transforming the FID converts this time-domain oscillation into the familiar spectrum of peaks, whose positions and splittings encode connectivity and conformation~\cite{bloch1946nuclear, hahn1950spin}.

To write the problem in a tractable form, NMR spectroscopists work in the \emph{rotating frame}: a reference frame that co-rotates with the spins at the spectrometer frequency $\nu_0 = \gamma B_0/(2\pi)$ ($\gamma$ is the proton gyromagnetic ratio, $B_0$ the static field). In this frame, the fast Larmor precession is removed, leaving only the slow, chemically informative differential precession between protons. The effective spin Hamiltonian is then~\cite{levitt2008spin, abragam1961principles, slichter1990principles}:
\begin{equation}
    \hat{H} = \sum_{i=1}^{N} \omega_i \hat{I}_{z,i} + 2\pi\sum_{i<j} J_{ij}\, \hat{\mathbf{I}}_i \cdot \hat{\mathbf{I}}_j,
    \label{eq:hamiltonian}
\end{equation}
where $\hat{\mathbf{I}}_i = (\hat{I}_{x,i}, \hat{I}_{y,i}, \hat{I}_{z,i})$ are the spin-$\frac{1}{2}$ operators of proton $i$ (eigenvalues $\pm\tfrac{1}{2}$, in units of $\hbar$). The first term shifts each resonance to angular frequency $\omega_i = 2\pi\nu_0(\delta_i - \delta_{\mathrm{ref}}) \times 10^{-6}$ relative to the reference compound. The isotropic scalar coupling $2\pi J_{ij}\,\hat{\mathbf{I}}_i \cdot \hat{\mathbf{I}}_j$ (in rad/s, with $J_{ij}$ reported in Hz) is the quantum-mechanical origin of multiplet splitting; its isotropic dot-product form holds for rapidly tumbling molecules in solution, where anisotropic dipolar couplings average to zero. Throughout this work we use $B_0 = \SI{11.74}{\tesla}$, giving $\nu_0 = \SI{500}{\mega\hertz}$.

After a $90^{\circ}_y$ pulse, the spin ensemble starts with all magnetization along $x$. At high temperature the relevant part of the density matrix is proportional to $\hat{I}_x^{\mathrm{tot}}$, and its time evolution gives the same spectral information as propagating the pure product state $\ket{\psi_0} = \ket{+_x}^{\otimes N}$, with each proton initialized in the $+x$ eigenstate, up to an overall normalization~\cite{levitt2008spin, abragam1961principles}. The FID signal is the complex transverse magnetization:
\begin{equation}
    S(t) = \langle \hat{I}_x^{\mathrm{tot}}(t) \rangle + i\, \langle \hat{I}_y^{\mathrm{tot}}(t) \rangle,
    \label{eq:fid}
\end{equation}
with $\ket{\Psi(t)} = e^{-i\hat{H}t}\ket{\psi_0}$ evolved under $\hat{H}$. The spectrum follows from a Fourier transform with an exponential window~\cite{kubo1957statistical, ernst1987principles}:
\begin{equation}
    S(\omega) = \int_0^T S(t)\, e^{-\lambda t}\, e^{-i\omega t}\, dt.
    \label{eq:spectrum}
\end{equation}
The apodization factor $e^{-\lambda t}$ mimics the $T_2^*$ decay that broadens experimental lines (we use $\lambda = \SI{3.0}{\per\second}$, giving $\sim\SI{1}{\hertz}$ Lorentzian linewidth) and damps ringing from the finite acquisition window $T$. The computational task is therefore to propagate $\ket{\Psi(t)}$ faithfully over $T = \SI{1.0}{\second}$ and evaluate $S(t)$ at each timestep: a quantum dynamics problem whose cost grows as $2^N$ with the number of coupled protons $N$.

\subsection{Interaction-Frame Propagation}
\label{sec:interaction}

The rotating-frame Hamiltonian of Eq.~\eqref{eq:hamiltonian} contains two terms separated by orders of magnitude in energy. The one-body chemical-shift term $\hat{H}_0 = \sum_i \omega_i \hat{I}_{z,i}$ has matrix elements of order $10^2$--$10^3$\,rad/s, while the scalar coupling $\hat{H}_1 = 2\pi\sum_{i<j}J_{ij}\,\hat{\mathbf{I}}_i\cdot\hat{\mathbf{I}}_j$ is of order $10^1$\,rad/s. An explicit integrator must resolve the fastest oscillation present in the propagated state, so the chemical-shift precession (though spectroscopically a trivial single-spin phase $e^{-i\omega_i t}$) sets the usable timestep and is responsible for the steep growth in step count reported in Table~\ref{tab:scaling}. This is numerical \emph{stiffness} in the standard sense: the step size is dictated by the fastest frequency present rather than by the slower dynamics of interest. We quantify it by the offset-to-coupling ratio $\max_{ij}|\omega_i - \omega_j|/2\pi J_{ij}$.

This stiffness can be removed analytically by propagating in the \emph{interaction frame} of $\hat{H}_0$. Writing $\hat{U}_0(t) = e^{i\hat{H}_0 t}$ and $\ket{\Psi_I(t)} = \hat{U}_0(t)\ket{\Psi(t)}$, the transformed state obeys $i\partial_t\ket{\Psi_I} = \hat{H}_I(t)\ket{\Psi_I}$ with a generator that retains only the coupling, dressed by phases oscillating at the chemical-shift \emph{differences}:
\begin{equation}
  \hat{H}_I(t) = 2\pi\sum_{i<j}J_{ij}\Big[\hat{I}_{z,i}\hat{I}_{z,j}
  + \tfrac12\big(e^{i(\omega_i-\omega_j)t}\,\hat{I}_+^{(i)}\hat{I}_-^{(j)} + \mathrm{h.c.}\big)\Big].
  \label{eq:HI}
\end{equation}
The large one-body term has vanished from the generator; what remains has magnitude $\sim 2\pi J_{ij}$, the slow coupling scale. The price is that $\hat{H}_I(t)$ is explicitly time dependent, with off-diagonal elements oscillating at the chemical-shift differences. This is a good trade: the rate at which the state changes is bounded by the norm of the generator, not by how rapidly its phases oscillate, so $\ket{\Psi_I(t)}$ moves on the slow coupling timescale and the integrator can take correspondingly larger steps.

The transformation is the standard interaction (Dirac) picture of quantum mechanics and involves no approximation: $\ket{\Psi(t)} = \hat{U}_0^\dagger(t)\ket{\Psi_I(t)}$ reproduces the rotating-frame state exactly at every instant. In particular, the chemical shifts are not discarded; they are applied analytically rather than integrated numerically. The laboratory-frame signal follows by transforming back,
\begin{equation}
  \langle\hat{I}_+^{\mathrm{tot}}(t)\rangle = \sum_k e^{i\omega_k t}\,\bra{\Psi_I}\hat{I}_+^{(k)}\ket{\Psi_I}.
  \label{eq:backtransform}
\end{equation}
Each factor $e^{i\omega_k t}$ is precisely the offset precession of spin $k$; in the frequency domain it translates the multiplet of that spin, which the interaction-frame dynamics generates around zero frequency, to its chemical-shift position $\omega_k$. Peak positions are thus restored exactly by the analytic phases, while the multiplet structure is carried by the propagated coupling dynamics. The resulting spectrum is identical to the rotating-frame one; only the numerical cost differs, because the integrator no longer resolves single-spin precession that is known in closed form.

\subsection{Molecular Spin Systems}

We study four molecules of increasing spectral complexity (Fig.~\ref{fig:molecules}), chosen to cover a range of coupling topologies representative of common organic structural motifs:
\begin{itemize}
    \item \textbf{2-spin:} \textit{cis}-3-Chloroacrylic acid. The two vinyl protons are coupled by a single vicinal $^3J$ constant, producing a simple doublet, the minimal non-trivial NMR pattern.
    \item \textbf{3-spin:} 1,2,4-Trichlorobenzene. Three aromatic protons at the 3-, 5-, and 6-positions are magnetically inequivalent and carry different $J$ couplings, giving an ABX-type multiplet structure.
    \item \textbf{4-spin:} 3-Methyleneoxetan-2-one. The exocyclic methylene and ring protons form a tightly coupled four-spin system with several overlapping multiplets.
    \item \textbf{5-spin:} Nitrobenzene. The five aromatic protons split into ortho, meta, and para sets by molecular symmetry; their coupling pattern is a textbook example of monosubstituted-benzene multiplet structure.
\end{itemize}
Hamiltonian parameters (chemical shifts and $J$-couplings) are taken from experimental databases~\cite{dashti2017applications, hqs2023}. For each system, the exact reference spectrum is computed by dense matrix exponentiation of $\hat{H}$. For the interaction-frame comparison (Table~\ref{tab:interaction}) we additionally include \textbf{sucrose} ($N = 14$), a disaccharide with fourteen inequivalent protons. Its $J$-coupling network splits into three disconnected groups of 7, 5, and 2 protons, so it is a comparatively benign large system rather than a strongly correlated one; we quantify this in Sec.~\ref{sec:interaction_results}. Its exact reference is obtained by sparse Krylov propagation ($\texttt{expm\_multiply}$) rather than dense diagonalization, and the NQS never enumerates its $2^{14}$-dimensional Hilbert space.

\section{Computational Methods}

\subsection{Neural Quantum State Representation}

We represent the instantaneous quantum state using a Restricted Boltzmann Machine (RBM), a two-layer neural network comprising visible units (corresponding to the $N$ physical spins) and $M = \alpha N$ hidden units. The wavefunction amplitude for a spin configuration $\mathbf{s} = (s_1, \ldots, s_N) \in \{-1, +1\}^N$ takes the form~\cite{carleo2017solving, becca2017quantum}:
\begin{equation}
    \psi(\mathbf{s}) = e^{\sum_i a_i s_i} \prod_{j=1}^{M} 2\cosh\left(b_j + \sum_i W_{ji} s_i\right),
    \label{eq:rbm}
\end{equation}
where $\{a_i\} \in \mathbb{C}^N$, $\{b_j\} \in \mathbb{C}^M$, and $W \in \mathbb{C}^{M \times N}$ are variational parameters. The hidden unit ratio $\alpha$ controls the network's expressive capacity.

The RBM is attractive here because the hidden units can be summed out analytically, giving $\psi(\mathbf{s})$ in $\mathcal{O}(\alpha N^2)$ operations. Complex-valued parameters are needed to capture the phases that develop during unitary evolution~\cite{schmitt2020quantum}. For states with limited entanglement, the required number of hidden units scales polynomially~\cite{gao2017efficient}, but volume-law entangled states may demand exponentially many. It is not known \textit{a priori} which regime NMR time evolution falls into.

\subsection{Time-Dependent Variational Principle}

Exact time evolution under $\hat{H}$ generates a trajectory $\ket{\Psi(t)} = e^{-i\hat{H}t}\ket{\Psi(0)}$ through the full $2^N$-dimensional Hilbert space. The NQS ansatz restricts the state to the variational manifold $\mathcal{M} = \{\ket{\Psi(\boldsymbol{\theta})} : \boldsymbol{\theta} \in \mathbb{C}^P\}$, and TDVP~\cite{mclachlan1964variational, haegeman2011time, yuan2019theory} provides the optimal equations of motion for the parameters $\boldsymbol{\theta}(t)$ on this manifold.

We project the Schrödinger equation $\partial_t\ket{\Psi} = -i\hat{H}\ket{\Psi}$ onto the tangent space $T_{\boldsymbol{\theta}}\mathcal{M}$, spanned by $\{\ket{\partial_\mu\Psi}\}$. The time derivatives $\dot{\theta}_\mu$ are chosen so that the resulting state update, $\partial_t\ket{\Psi(\boldsymbol{\theta})} = \sum_\mu \dot{\theta}_\mu\ket{\partial_\mu\Psi}$, is as close as possible, in the Hilbert-space norm, to the exact right-hand side $-i\hat{H}\ket{\Psi}$. This least-squares condition reads:
\begin{equation}
    \min_{\dot{\boldsymbol{\theta}}} \left\| \sum_\mu \dot{\theta}_\mu \ket{\partial_\mu \Psi} + i\hat{H}\ket{\Psi} \right\|^2,
    \label{eq:tdvp}
\end{equation}
whose solution is the linear system of equations of motion~\cite{schmitt2020quantum}:
\begin{equation}
    \QGT\,\dot{\boldsymbol{\theta}} = -i\,\mathbf{F}.
    \label{eq:eom}
\end{equation}

Both $\QGT$ and $\mathbf{F}$ are estimated stochastically from $n_{\mathrm{samples}}$ configurations drawn from the Born distribution $|\psi(\mathbf{s})|^2$. Defining the log-derivative $O_\mu(\mathbf{s}) = \partial_{\theta_\mu} \ln \psi(\mathbf{s})$ and the local energy $E_{\mathrm{loc}}(\mathbf{s}) = \sum_{\mathbf{s}'} \langle \mathbf{s}|\hat{H}|\mathbf{s}'\rangle \psi(\mathbf{s}')/\psi(\mathbf{s})$, the quantum geometric tensor (quantum Fisher information matrix)~\cite{sorella1998green, stokes2020quantum} and the force vector are the connected covariances:
\begin{align}
    S_{\mu\nu} &= \langle O_\mu^* O_\nu \rangle_c, \label{eq:qgt}\\
    F_\mu &= \langle O_\mu^* E_{\mathrm{loc}} \rangle_c, \label{eq:force}
\end{align}
where $\langle A B \rangle_c \equiv \langle AB\rangle - \langle A\rangle\langle B\rangle$ denotes the connected expectation value. $S_{\mu\nu}$ measures how independently different parameters deform the quantum state (it is the Gram matrix of the tangent vectors in the Fubini--Study metric); $\mathbf{F}$ measures how much each parameter direction overlaps with the exact Schrödinger drive $-i\hat{H}\ket{\Psi}$. These equations of motion are reparameterization-covariant: the physical trajectory is independent of how the network is parameterized, and the method reduces to natural gradient descent in the geometry of quantum state space~\cite{stokes2020quantum, hackl2020geometry}.

\subsection{Projected Time-Dependent Variational Monte Carlo}

The projected time-dependent Variational Monte Carlo (p-tVMC) method~\cite{sinibaldi2023unbiasing} takes a different approach from TDVP: rather than integrating a differential equation on $\mathcal{M}$, it advances the state by \emph{projecting} the exactly evolved quantum state back onto the variational manifold at each discrete timestep. Concretely, after applying the exact propagator for one step, the target state is
\begin{equation}
    \ket{\Phi_t} = e^{-i\hat{H}\Delta t} \ket{\Psi(\boldsymbol{\theta}_t)},
    \label{eq:target_state}
\end{equation}
which in general lies outside $\mathcal{M}$. The projection finds the variational state $\ket{\Psi(\boldsymbol{\theta}_{t+1})} \in \mathcal{M}$ closest to $\ket{\Phi_t}$ by minimizing the quantum infidelity:
\begin{equation}
    \mathcal{I}(\boldsymbol{\theta}) = 1 - \frac{|\braket{\Psi(\boldsymbol{\theta})}{\Phi_t}|^2}{\braket{\Psi(\boldsymbol{\theta})}{\Psi(\boldsymbol{\theta})} \braket{\Phi_t}{\Phi_t}}.
    \label{eq:infidelity}
\end{equation}
$\mathcal{I} = 0$ means perfect overlap (the variational ansatz can represent the propagated state exactly); $\mathcal{I} = 1$ means the two states are orthogonal.

In our implementation, the exact propagator $e^{-i\hat{H}\Delta t}$ is precomputed once as a dense $2^N \times 2^N$ matrix via matrix exponentiation and cached throughout the simulation; this is feasible for $N \leq 12$ and eliminates propagator discretization errors from the benchmark. The infidelity $\mathcal{I}$ and its gradient are estimated by Monte Carlo: configurations $\boldsymbol{\sigma}$ are sampled from $|\psi_{\boldsymbol{\theta}}(\boldsymbol{\sigma})|^2$ and independently $\boldsymbol{\eta}$ from $|\phi_t(\boldsymbol{\eta})|^2$. The key estimator is
\begin{equation}
    I_{\mathrm{loc}}(\boldsymbol{\sigma}, \boldsymbol{\eta}) = \frac{\phi_t(\boldsymbol{\sigma})}{\psi_{\boldsymbol{\theta}}(\boldsymbol{\sigma})} \cdot \frac{\psi_{\boldsymbol{\theta}}(\boldsymbol{\eta})}{\phi_t(\boldsymbol{\eta})},\quad
    \mathcal{I} = 1 - |\langle I_{\mathrm{loc}} \rangle|^2,
\end{equation}
where each factor is the amplitude ratio of the two wavefunctions at a sampled configuration; together they form an importance-sampled estimator of the squared overlap $|\langle\Psi(\boldsymbol{\theta})|\Phi_t\rangle|^2$.
Gradient variance is reduced by the control variate~\cite{sinibaldi2023unbiasing}
\begin{equation}
    I_{\mathrm{loc}}^{\mathrm{CV}} = \mathrm{Re}\{I_{\mathrm{loc}}\} - c\!\left(|1 - I_{\mathrm{loc}}|^2 - 1\right),
    \label{eq:cv}
\end{equation}
with $c = -1/2$ optimal for small infidelities. Gradients are computed by automatic differentiation and passed to the Adam optimizer~\cite{kingma2015adam}; optionally, Stochastic Reconfiguration (SR) preconditioning can be applied to the updates.

Compared with TDVP, p-tVMC sidesteps QGT inversion, and the per-step $\mathcal{I}$ gives a running diagnostic of how well the variational manifold tracks the true evolution. The per-step cost $\mathcal{O}(n_{\mathrm{iter}} \cdot n_{\mathrm{samples}} \cdot P)$ exceeds the $\mathcal{O}(n_{\mathrm{samples}} \cdot P^2)$ of TDVP, but iterative optimization avoids the regularization bias that accumulates in TDVP. For $N \gtrsim 12$, the dense propagator must be replaced by a Trotter--Suzuki decomposition; this introduces an $\mathcal{O}(\Delta t^2)$ Trotter error per step and is the main scalability bottleneck of the current implementation.

\subsection{Implementation}
\label{sec:implementation}

We implement both the NQS-TDVP and p-tVMC methods using NetKet~\cite{vicentini2022netket}, an open-source library built on JAX that provides automatic differentiation, just-in-time compilation, and GPU acceleration. The variational state employs Metropolis-Hastings sampling with local spin-flip proposals, using $n_{\mathrm{samples}} = 4096$ configurations per time step. This is admittedly excessive for the $N \le 5$ systems, where exact summation over all basis configurations is possible; we keep Monte Carlo sampling rather than exact summation throughout, to stay true to the larger systems for which the method is meant, where summing over all configurations is out of reach. All benchmarks were run on a single CPU node using 8 threads (Intel Xeon, \SI{2.4}{\giga\hertz}); the wall times reported in Table~\ref{tab:scaling} refer to this configuration. For direct method comparison, both TDVP and p-tVMC use the same hidden unit ratio $\alpha = 4$. For the TDVP-only scaling results in Table~\ref{tab:scaling}, we additionally identified $\alpha$ that minimizes the spectral MSE for each system. The MSE is computed over the spectral region containing peaks after normalizing both the NQS and the exact spectra to the unit maximum.

\textit{TDVP integration:} Time propagation uses the Heun method (second-order Runge-Kutta) with timestep $\Delta t$, and the linear system $\QGT \cdot \Delta\boldsymbol{\theta} = -i\,\mathbf{F}\,\Delta t$ is solved via singular value decomposition with diagonal regularization $\epsilon = 0.01$ for numerical stability. For each system we vary $\Delta t$ over four to five values, keeping the total simulation time fixed at $T = \SI{1.0}{\second}$, and use the value that minimizes the spectral MSE; the scanned ranges run from $0.1$--$0.005$\,s for the 2-spin system down to $10^{-3}$--$10^{-4}$\,s for the 5-spin system, and the optima are listed in Table~\ref{tab:scaling}.

\textit{p-tVMC optimization:} The exact propagator $U(\Delta t) = e^{-i\hat{H}\Delta t}$ is precomputed once as a dense $2^N \times 2^N$ matrix via matrix exponentiation and cached throughout the simulation. At each time step, the infidelity $\mathcal{I}$ is minimized using the Adam optimizer~\cite{kingma2015adam} with learning rate $\eta = 0.01$ and $n_{\mathrm{iter}} = 50$ gradient updates per step. The control variate coefficient is $c = -0.5$. An optional Stochastic Reconfiguration preconditioner (diagonal shift $\epsilon = 0.01$) was tested but not found to improve results consistently.

The RBM parameters are initialized with small random values (scale $= 0.01$, seed $= 42$). Because the product state $\ket{+_x}^{\otimes N}$ can be represented exactly by an RBM, convergence to the correct initial state is verified by checking that $\langle \hat{I}_x^{\mathrm{tot}} \rangle = N/2$ to within statistical error before beginning time propagation. Molecular Hamiltonian parameters (chemical shifts and $J$-couplings) are taken from experimental databases~\cite{dashti2017applications,hqs2023}.

\section{Results}

\subsection{Rotating-Frame Benchmarks}

\begin{figure}[!ht]
    \centering
    \includegraphics[width=\columnwidth]{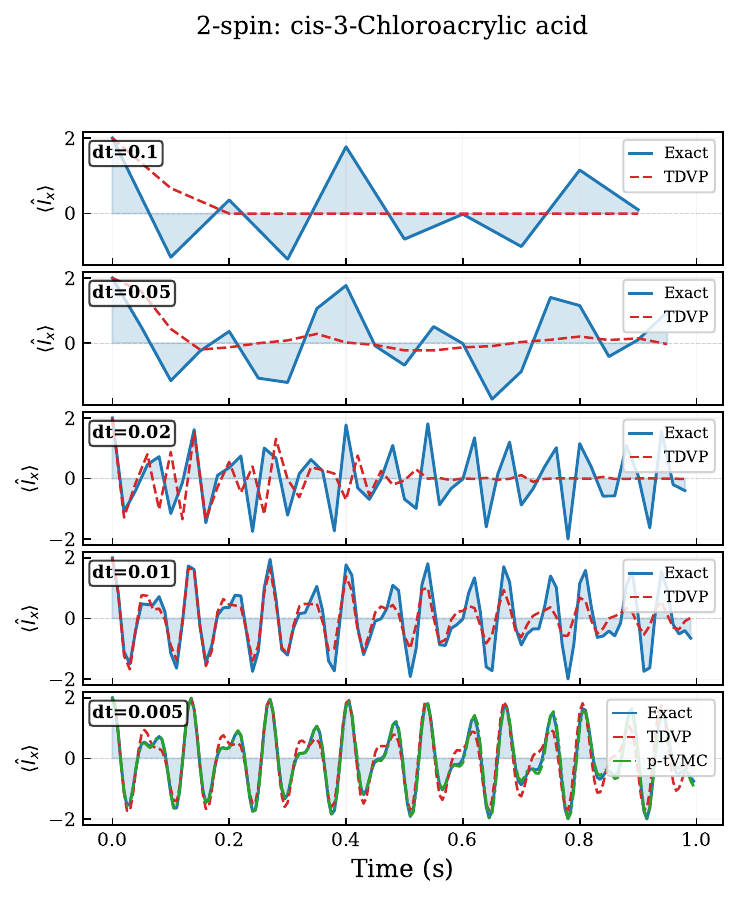}
    \caption{Transverse magnetization $\langle \hat{I}_x(t) \rangle$ for the 2-spin system at decreasing $\Delta t$ (coarser at top, finest at bottom). Blue shaded: exact evolution; red dashed: NQS-TDVP; green dash-dot: p-tVMC (finest panel only). Both methods converge to the exact trajectory as $\Delta t$ decreases.}
    \label{fig:dynamics}
\end{figure}

We begin with the 2-spin system, where the dynamics can be visualized directly. Figure~\ref{fig:dynamics} shows the transverse magnetization $\langle \hat{I}_x(t) \rangle$ over the full $T = \SI{1.0}{\second}$ window at successive integration timesteps. The beating pattern is the time-domain signature of the doublet: two precessing frequencies that alternately reinforce and cancel. At coarse $\Delta t = 0.1$\,s (10 steps) both TDVP and p-tVMC drift from the exact solution in amplitude and phase. By $\Delta t = 0.005$\,s ($N_t = 200$) the TDVP curve is indistinguishable from the exact, and p-tVMC tracks it closely, though with larger residual error (quantified below). The exponential window $e^{-\lambda t}$ downweights the late-time signal, where errors accumulate most, so the spectral accuracy is somewhat better than the raw time-domain discrepancy suggests.

\begin{figure*}[!ht]
    \centering
    \begin{tabular}{cc}
        \includegraphics[width=0.48\textwidth]{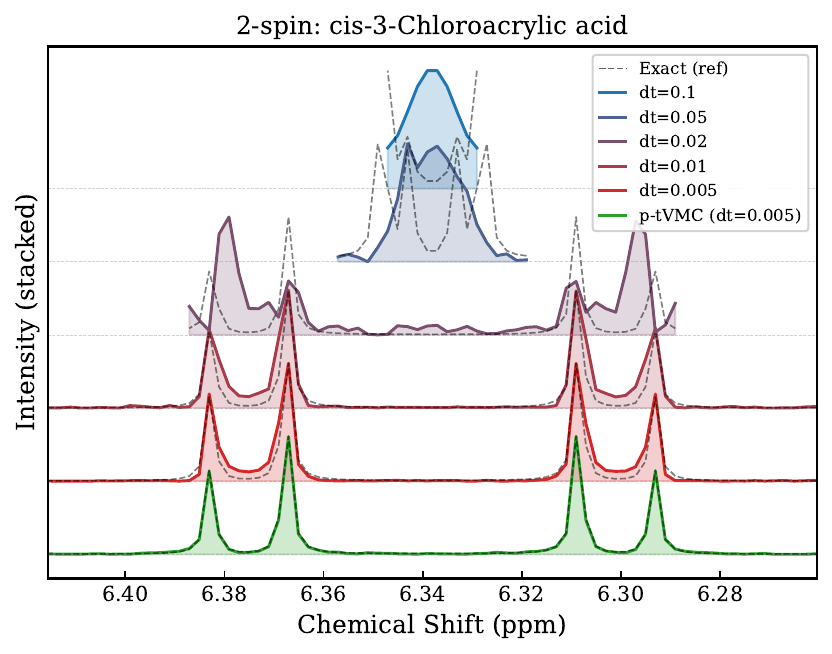} &
        \includegraphics[width=0.48\textwidth]{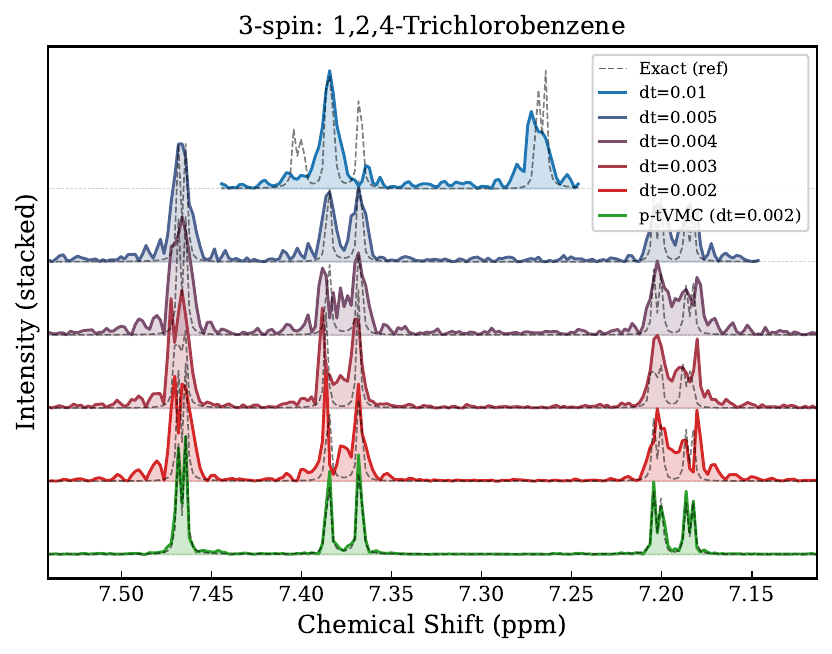} \\[4pt]
        \includegraphics[width=0.48\textwidth]{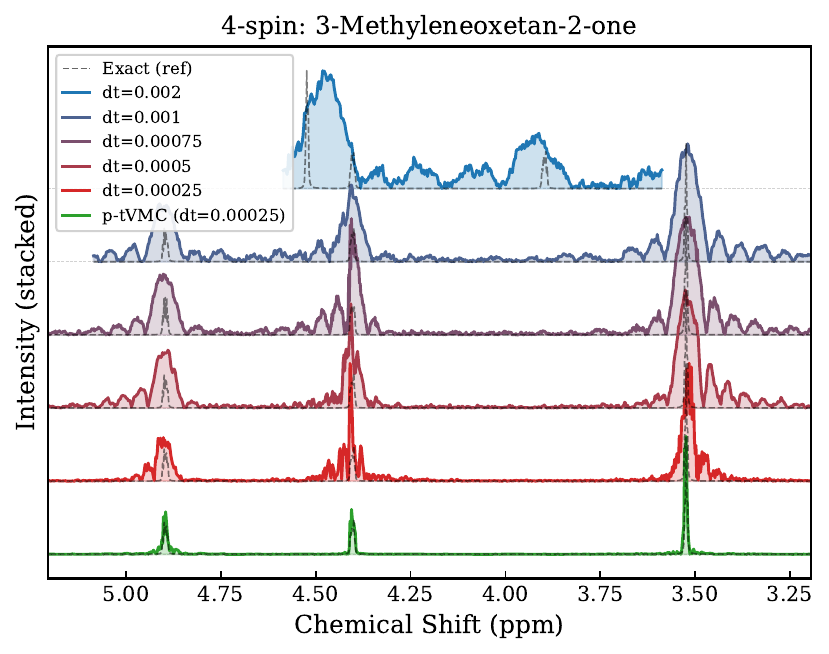} &
        \includegraphics[width=0.48\textwidth]{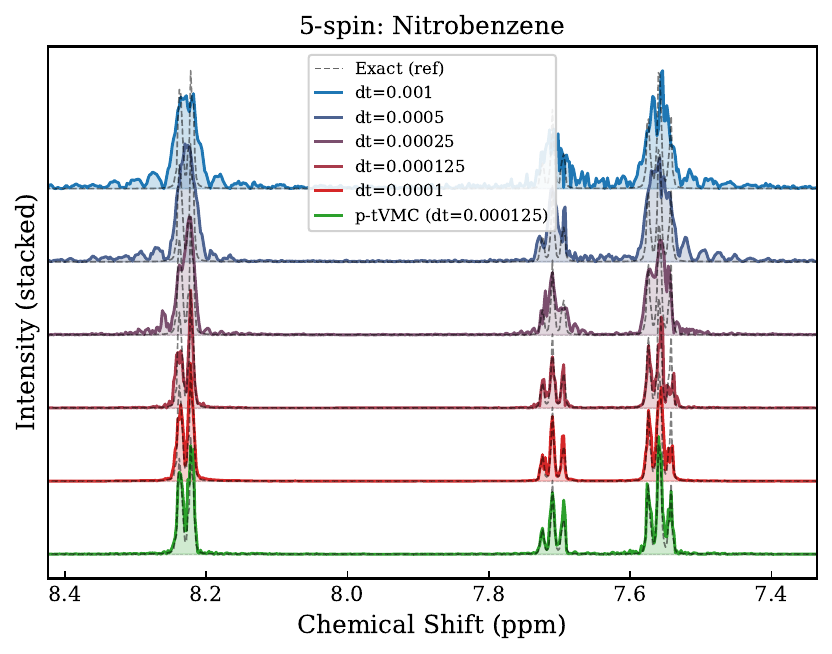}
    \end{tabular}
    \caption{NMR spectra for all four molecular spin systems, stacked by integration timestep $\Delta t$ (coarsest at top, finest at bottom). Colored fills: NQS-TDVP; green: p-tVMC at the finest timestep; dashed black: exact reference. The exact reference is evaluated on the same time grid as the NQS run in each row ($N_t$ points at the corresponding $\Delta t$) and Fourier-transformed identically, so its spectral resolution improves down the stack alongside the NQS spectra; rows differ in resolution, not in the underlying spectrum. All four systems show progressive convergence to the exact spectrum as $\Delta t$ decreases; p-tVMC reproduces the same spectral features at the finest level, though with larger residual error than TDVP (see text).}
    \label{fig:spectra}
\end{figure*}

The spectral consequences are shown in Fig.~\ref{fig:spectra} for all four molecules. Each panel stacks the NQS spectra at successive $\Delta t$ values (coarsest at top, finest at bottom), with p-tVMC in green and the exact reference as dashed black. For the 2-spin doublet of \textit{cis}-3-chloroacrylic acid, line positions are resolved already at $N_t = 20$ steps; the fine splitting and clean baseline emerge only at $N_t = 200$. The ABX multiplet of 1,2,4-trichlorobenzene (6.8--7.8 ppm) sharpens progressively with smaller $\Delta t$, though low-intensity oscillations between peaks persist at intermediate step counts. The four-spin system shows noisier baselines, consistent with the higher initialization sensitivity discussed below, but reproduces the correct multiplet pattern at fine $\Delta t$. Nitrobenzene, with its ortho/meta/para sub-patterns, requires $N_t = 10{,}000$ steps, fifty times more than the 2-spin case: much finer spectral structure must be resolved over the same 1\,s window. In every case, the simulated spectra converge toward the exact reference as $\Delta t$ decreases.

\begin{figure*}[!ht]
    \centering
    \includegraphics[width=\textwidth]{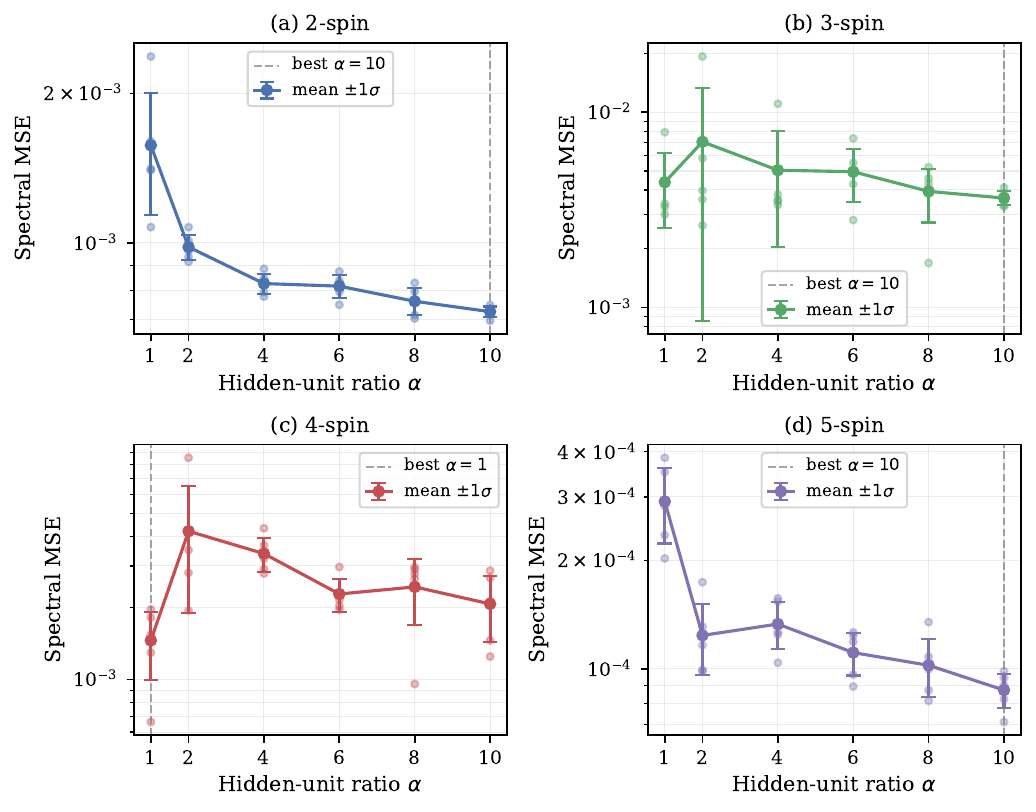}
    \caption{Spectral MSE as a function of the hidden-unit ratio $\alpha$ for each molecular spin system (5 independent random seeds per point). Filled circles show individual seeds; connected markers show mean $\pm 1\sigma$; dashed vertical lines mark the optimal $\alpha$. The 4-spin molecule is anomalous: it achieves its best accuracy at $\alpha = 1$, whereas all other systems benefit from larger networks.}
    \label{fig:alpha_sweep}
\end{figure*}

Table~\ref{tab:scaling} and Fig.~\ref{fig:alpha_sweep} quantify the computational cost across all four systems (MSE averaged over 5 random seeds). The required $\Delta t$ shrinks with $N$, tracking the growing spectral bandwidth of larger spin systems. MSE does not increase monotonically: the 4-spin molecule achieves lower mean error than 3-spin, and the 5-spin the lowest of all, reflecting genuine differences in spectral complexity between molecules. Initialization sensitivity tells a different story: the coefficient of variation grows from 3\% at $N = 2$ to 32\% at $N = 4$, so single-seed results can be misleading. In general, we expect spectral MSE to decrease as the network size increases, and this is reflected in the $2-$, $3-$, $5-$spin systems. For the 4-spin system, the best accuracy is at $\alpha = 1$. This anomaly could have arisen from chance, such as obtaining more favorable initialization seeds, or having specific coupling parameters. Repeating the sweep on a different 4-spin molecule would distinguish the two.

\begin{table}[!ht]
    \centering
    \caption{Scaling of TDVP computational requirements with system size for $T = \SI{1.0}{\second}$ total simulation time. $\Delta t_{\mathrm{opt}}$: optimal timestep; $N_t$: number of integration steps; $T_{\mathrm{wall}}$: mean wall-clock time per seed (8 CPU threads, Intel Xeon, \SI{2.4}{\giga\hertz}). MSE = mean $\pm$ std over 5 random seeds at $\alpha_{\mathrm{opt}}$.}
    \label{tab:scaling}
    \begin{tabular}{lcccccc}
        \toprule
        $N$ & Dim & $\Delta t_{\mathrm{opt}}$ (s) & $N_t$ & $\alpha_{\mathrm{opt}}$ & MSE & $T_{\mathrm{wall}}$ \\
        \midrule
        2 & 4 & 0.005 & 200 & 10 & $(7.3 \pm 0.2){\times}10^{-4}$ & $31$\,s \\
        3 & 8 & 0.002 & 500 & 10 & $(3.6 \pm 0.3){\times}10^{-3}$ & $\sim$2\,min \\
        4 & 16 & $2.5{\times}10^{-4}$ & 4000 & 1 & $(1.5 \pm 0.5){\times}10^{-3}$ & $\sim$8\,min \\
        5 & 32 & $1.0{\times}10^{-4}$ & 10000 & 10 & $(8.7 \pm 1.0){\times}10^{-5}$ & $\sim$90\,min \\
        \bottomrule
    \end{tabular}
\end{table}

\begin{figure}[!ht]
    \centering
    \includegraphics[width=\columnwidth]{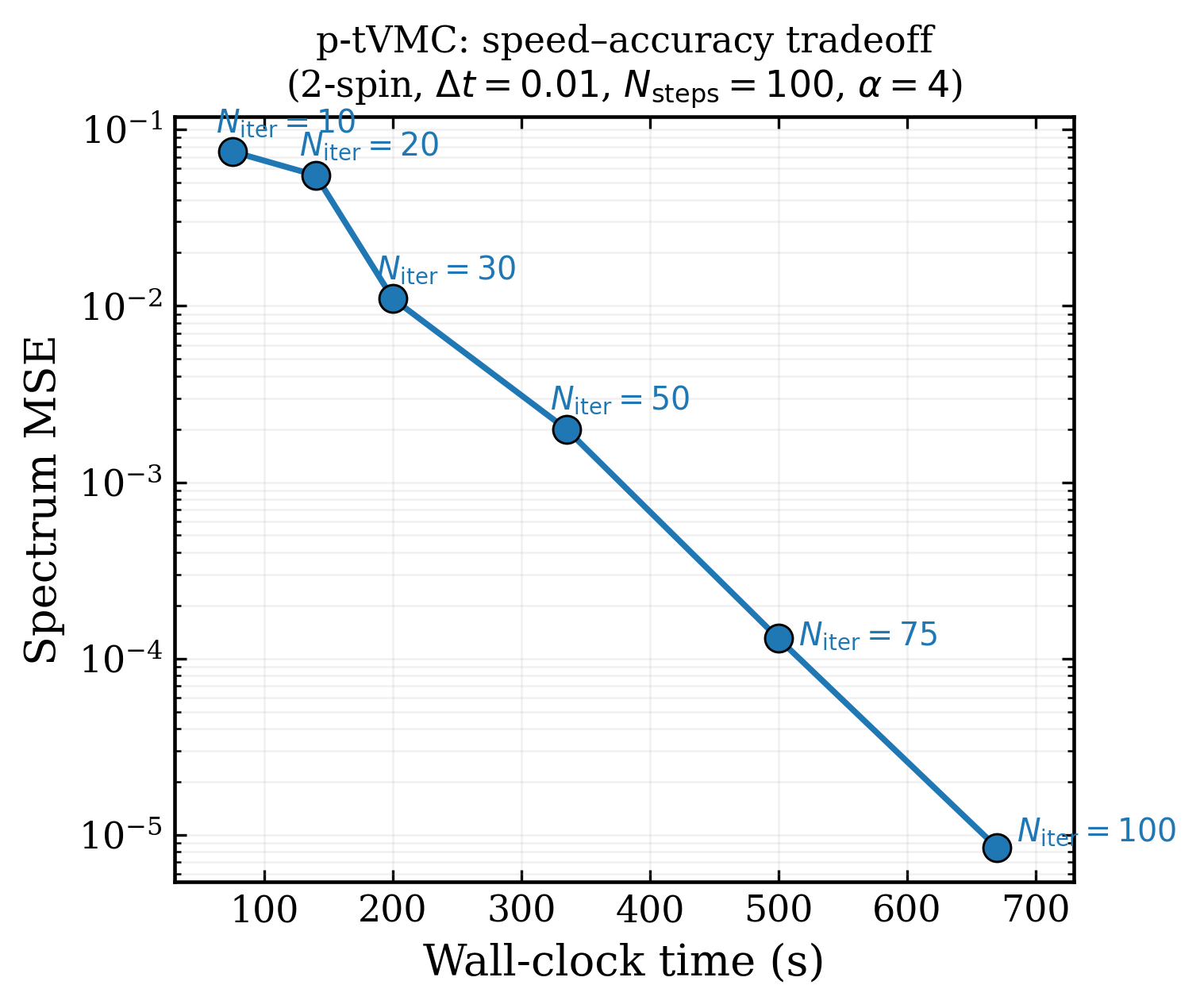}
    \caption{Speed--accuracy tradeoff for p-tVMC on the 2-spin system ($n_{\mathrm{iter}} \in \{10, 20, 30, 50, 75, 100\}$ gradient steps per timestep). Wall time scales linearly with $n_{\mathrm{iter}}$; MSE drops by four orders of magnitude, showing that accuracy is optimizer-limited rather than expressivity-limited at these system sizes.}
    \label{fig:ptvmc_tradeoff}
\end{figure}

The two methods handle the computation-accuracy tradeoff differently. TDVP solves the linear system $\QGT\,\dot{\boldsymbol{\theta}} = -i\mathbf{F}$ once per step at a cost that is fixed for given $n_{\mathrm{samples}}$ and $\alpha$. p-tVMC instead runs $n_{\mathrm{iter}}$ gradient steps per timestep ($n_{\mathrm{iter}} = 50$ for the spectra in Fig.~\ref{fig:spectra}, $\alpha = 4$). Across the four systems it reaches spectral MSEs between $2.6\times10^{-3}$ (5-spin, run at $n_{\mathrm{iter}} = 75$) and $3.7\times10^{-2}$ (3-spin), with per-step infidelities of $10^{-5}$--$10^{-2}$. These errors are higher than the TDVP optima in Table~\ref{tab:scaling}, though the comparison is not direct since TDVP used the molecule-specific optimal $\alpha$ while p-tVMC used $\alpha = 4$ throughout. Figure~\ref{fig:ptvmc_tradeoff} shows the iteration sweep for the 2-spin system: as $n_{\mathrm{iter}}$ increases from 10 to 100, wall time grows from ${\sim}80$\,s to ${\sim}670$\,s while MSE drops by nearly four orders of magnitude. The improvement is still monotonic at $n_{\mathrm{iter}} = 100$, confirming that accuracy at these system sizes is limited by optimizer budget rather than network expressivity.

\subsection{Effect of the Interaction Frame}
\label{sec:interaction_results}

\begin{figure}[!ht]
    \centering
    \includegraphics[width=\columnwidth]{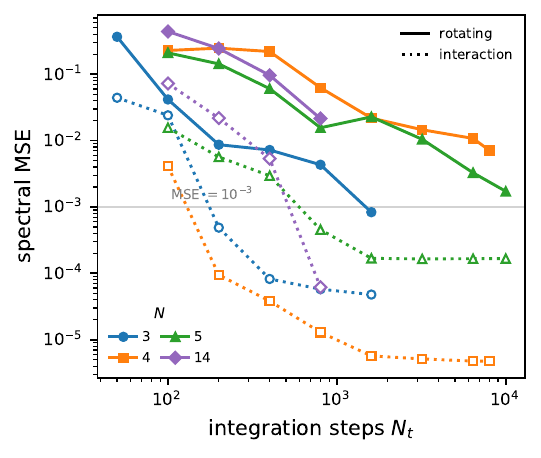}
    \caption{Spectral MSE versus number of integration steps $N_t$ for TDVP in the rotating frame (solid) and the interaction frame (dashed), for the three dense-coupling molecules ($N = 3, 4, 5$) and sucrose ($N = 14$). Curves are means over the available random seeds (up to three; the seed scatter is discussed in the text). The horizontal line marks MSE $=10^{-3}$. The rotating frame needs $\sim 10^3$--$10^4$ steps to approach this target; the interaction frame crosses it within a few hundred steps and then saturates at its error floor.}
    \label{fig:interaction_scaling}
\end{figure}

Most of the integration steps in Table~\ref{tab:scaling} are spent resolving the chemical-shift term rather than the couplings that carry the spectral information. Since the interaction-frame transformation of Eq.~\eqref{eq:HI} removes that term exactly, we used it to ask how much of the cost it accounts for. Because interaction frame transformation operator $\hat{U}_0(t)$ is diagonal and a product of single-spin $z$-rotations, applying it costs nothing and leaves the entanglement structure, and hence the RBM representability, of the state unchanged; for the ansatz of Eq.~\eqref{eq:rbm} it is absorbed \emph{exactly} into a drift of the complex visible biases, $a_i(t) = a_i(0) - i(\omega_i/2)\,t$. We repeated the TDVP propagation in both frames under otherwise identical settings: RBM with $\alpha = 4$ for the $N \le 5$ molecules and $\alpha = 2$ for the $N=14$ sucrose, $4096$ samples per step, and the same Heun integrator, QGT/SVD solver, and sampler. The rotating frame was pushed to $N_t$ as large as $10^4$ to follow its convergence. Figure~\ref{fig:interaction_scaling} shows the spectral MSE against the number of steps $N_t$ in both frames, and Table~\ref{tab:interaction} summarizes the contrast.

These runs are a controlled frame comparison at fixed network size, not accuracy-optimized, so their rotating-frame errors exceed the Table~\ref{tab:scaling} values at the same $N$, which used the molecule-specific $\alpha_{\mathrm{opt}}$ and timestep averaged over five seeds. Repeating the 4- and 5-spin comparison at $\alpha = 2$ leaves the picture unchanged (rotating frame at $1$--$2\times10^{-2}$ at $N_t = 3200$, interaction frame at its floor), so the contrast is not an artifact of the network size.

\begin{table*}[!ht]
    \centering
    \caption{Integration steps needed to reach spectral MSE $\le 10^{-3}$ in the rotating and interaction frames, under otherwise identical settings. ``Stiffness'' is the ratio $\max_{ij}|\omega_i-\omega_j|/2\pi J_{ij}$ of Sec.~\ref{sec:interaction}. A ``$>$'' entry means the threshold was not reached at the largest step count we ran, so the corresponding gain is a lower bound. ``Floor$_{\mathrm{int}}$'' is the interaction frame's error floor (the $N_t$-independent level at which its error saturates; see text); for $N=2$, where the two frames are identical, we quote the error at the largest step count run. Values are three-seed means where available; the seed scatter of the rotating frame is discussed in the text. Sucrose was run only to $N_t = 800$ in both frames and is a feasibility point rather than a converged comparison.}
    \label{tab:interaction}
    \begin{tabular}{lcccccc}
        \toprule
        & & & \multicolumn{2}{c}{$N_t$ to MSE $\le 10^{-3}$} & & \\
        \cmidrule(lr){4-5}
        Molecule & $N$ & Stiffness & rot & int & Gain & Floor$_{\mathrm{int}}$ \\
        \midrule
        \textit{cis}-chloroacrylic acid & 2  & 4.7  & 200          & 200          & $1\times$          & $1.2{\times}10^{-4}$ \\
        1,2,4-trichlorobenzene          & 3  & 15.5 & ${\sim}1600$ & ${\sim}200$  & ${\sim}8\times$    & $4.8{\times}10^{-5}$ \\
        3-methyleneoxetan-2-one         & 4  & 36.9 & $>8000$      & ${\sim}200$  & $\gtrsim 40\times$ & $5.0{\times}10^{-6}$ \\
        nitrobenzene                    & 5  & 39.9 & $>10^4$      & ${\sim}800$  & $\gtrsim 12\times$ & $1.7{\times}10^{-4}$ \\
        sucrose                         & 14 & 72.5 & $>800$       & $\le 800$    & --                 & $6.2{\times}10^{-5}$ \\
        \bottomrule
    \end{tabular}
\end{table*}

The size of the effect depends strongly on the molecule. For the 2-spin system, the two frames are equivalent, as expected: the chemical-shift evolution is a time-linear shift of the RBM visible biases (Sec.~\ref{sec:interaction}), which the integrator reproduces exactly, so there is nothing to remove. For the 3-spin molecule, where the rotating frame still reaches the target accuracy, the interaction frame attains MSE $< 10^{-3}$ in about $200$ steps rather than roughly $1600$ (three seeds; one remains marginally above $10^{-3}$ at $N_t = 1600$), an approximately eightfold reduction at fixed accuracy.

For the 4- and 5-spin molecules, the contrast is larger. The interaction frame drops below MSE $10^{-3}$ within a few hundred steps ($N_t \approx 200$ and $\approx 800$, respectively) and then flattens at an $N_t$-independent residual of $5\times10^{-6}$ and $1.7\times10^{-4}$, respectively: an \emph{error floor} set by sampling noise and the QGT regularization rather than by the time step. The rotating frame instead needs of order $10^3$--$10^4$ steps: pushed to the largest step counts we ran, its error keeps decreasing, reaching $7.1\times10^{-3}$ at $N_t = 8000$ (4-spin, convergence-run seed) and $(1.7\pm0.4)\times10^{-3}$ at $N_t = 10^4$ (5-spin, three seeds) (Fig.~\ref{fig:interaction_scaling}). This points to slow stiffness rather than a hard error floor: the rotating frame converges toward the same spectrum, but the large chemical-shift term forces a small time step. Individual seeds are not monotonic, however. At intermediate $N_t$ the rotating-frame error scatters strongly across seeds and reruns (at $N_t = 3200$ for the 4-spin system, four runs span $1.2$--$3.1\times10^{-2}$); we read this as sampling noise riding on a slowly decreasing signal, not as a plateau.

The advantage survives in wall-clock time. For the 4- and 5-spin systems, the interaction frame, whose dressed coupling generator is rebuilt at every step, costs $10$--$40\%$ more per step. The 4-spin system nevertheless reached MSE $\approx 5\times10^{-6}$ in ${\approx}\,\SI{450}{\second}$ ($N_t = 1600$), whereas the rotating frame was still at $7\times10^{-3}$ after ${\approx}\,\SI{2050}{\second}$ ($N_t = 8000$). Removing the large one-body term also reduces the magnitude of the force $\mathbf{F} = \langle O^* E_{\mathrm{loc}}\rangle$, and with it the bias from the QGT diagonal shift; consistent with this, the interaction-frame results were markedly more reproducible across seeds (spreads below $10\%$) than the rotating-frame ones.

Sucrose ($N = 14$) probes a larger system, sampled by Monte Carlo rather than full enumeration. Within the $N_t \le 800$ steps we ran, the interaction frame reaches MSE ${\sim}6\times10^{-5}$ while the rotating frame is still at ${\sim}2\times10^{-2}$ (one of three seeds reached $4\times10^{-3}$); we did not push the rotating frame to convergence and quote sucrose only as a feasibility point.

\begin{figure}[!ht]
    \centering
    \includegraphics[width=\columnwidth]{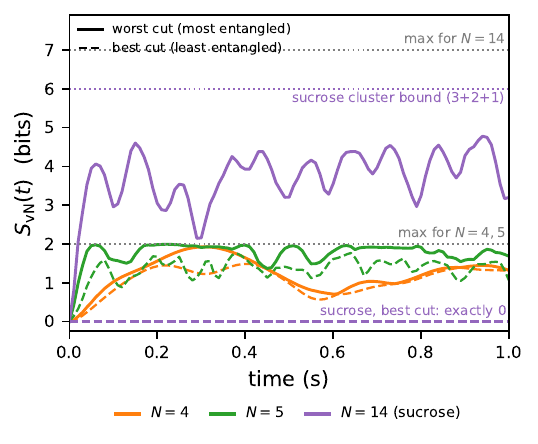}
    \caption{Bipartite von Neumann entanglement entropy of the exact state during the evolution, maximized (solid, ``worst cut'') and minimized (dashed, ``best cut'') over balanced bipartitions, for the 4- and 5-spin molecules and sucrose ($N = 14$). Dotted gray lines mark the maximal possible values for arbitrary states; the dotted purple line marks sucrose's cluster bound of $3+2+1 = 6$ bits. For sucrose the minimum is exactly zero at all times: its coupling graph consists of three disconnected clusters (7, 5, and 2 protons) and the state stays a product over them, so a balanced $7\,|\,7$ cut can fall on a cluster boundary; a cut crossing all three clusters instead sums their contributions and reaches $4.8$ bits (bounded by $3+2+1=6$, not $7$), while the entanglement \emph{within} the largest cluster reaches $2.9$ of a maximal $3$ bits. Computed by exact Krylov propagation and singular-value decomposition over all balanced bipartitions ($2|2$, $2|3$, and $7|7$ for $N = 4, 5, 14$).}
    \label{fig:entanglement}
\end{figure}

The successful simulation of the 14-proton molecule points works in favor of NQ. The successful run at $N = 14$ is concrete evidence that the approach extends well beyond the few-spin benchmarks. We further highlight that a larger molecule is not necessarily a harder one. Figure~\ref{fig:entanglement} shows the bipartite von Neumann entanglement entropy of the \emph{exact} state along the trajectory, maximized (``worst cut'') and minimized (``best cut'') over balanced bipartitions. The 4- and 5-spin molecules entangle across every balanced cut, with the worst approaching the maximal $2$ bits. Sucrose does not: its $J$-coupling network splits into three disconnected clusters (7, 5, and 2 protons), so the problem factorizes into independent blocks and becomes easier than its size suggests, its cost set by the largest block of 7 spins rather than by all 14. In this case, the coupling structure can further simplify molecules. More broadly, the appeal of NQS for larger systems is that, unlike tensor networks, they are not bounded by connectivity or entanglement across a cut~\cite{deng2017quantum}. They can remain efficient as that entanglement grows.

Two caveats bound this comparison. First, both frames used the same second-order integrator; a higher-order or adaptive scheme would also shorten the rotating-frame calculation, so the interaction frame is a cheap and exact way to handle the chemical-shift contribution, not the only one. Second, for the $N \le 5$ molecules, the $4096$ samples far exceed the Hilbert-space dimension ($\le 32$), so those comparisons are effectively at the exact-summation limit. Sucrose is the only genuine test of sparse stochastic sampling ($4096$ samples in a $2^{14} = 16384$-dimensional space); the low MSE reached there is an encouraging first data point rather than a systematic study of how accuracy scales with sample count, and the behaviour of larger, more strongly correlated systems, where entanglement rather than the chemical shift dominates the cost, remains open.

\section{Discussion}

\subsection{Method Comparison and Error Budget}

TDVP is cheaper per step and works well when the QGT stays well-conditioned; p-tVMC is more expensive but gives a per-step error diagnostic (the infidelity) and avoids QGT inversion entirely. At larger system sizes, the infidelity could serve as an adaptive convergence criterion, allocating fewer optimization iterations when the variational manifold tracks the true evolution closely and more when it does not. TDVP has no comparable built-in quality indicator, and its per-step cost is fixed regardless of local approximation quality.

The spectral error has four contributions, whose relative importance differs between the two methods. (i)~\textit{Integration/projection error.} The Heun integrator used in TDVP has $\mathcal{O}(\Delta t^3)$ local truncation error, giving $\mathcal{O}(T \Delta t^2)$ global error; higher-order integrators would reduce this. In p-tVMC, the analogous quantity is the residual infidelity per step. (ii)~\textit{Statistical noise.} Finite sampling ($n_{\mathrm{samples}} = 4096$) introduces fluctuations in the QGT and forces (TDVP) or the infidelity gradient (p-tVMC), with standard errors of ${\sim}\,1.5\%$ of mean values~\cite{chen2024efficient}. (iii)~\textit{QGT regularization (TDVP only).} The diagonal shift $\epsilon = 0.01$ biases the equations of motion; reducing $\epsilon$ helps but amplifies sampling noise. p-tVMC avoids this trade-off. (iv)~\textit{Variational expressivity.} If the true state develops entanglement that the RBM cannot efficiently represent, both methods will drift from the exact trajectory; this is the most fundamental limitation.
\subsection{Implications for Practical NMR Simulation}

For 2--5 spins, exact diagonalization is trivially faster (the largest Hilbert space here is $2^5 = 32$); the point of the benchmark is to characterize how NQS accuracy and cost scale before reaching sizes where exact methods fail. Exact propagation scales as $\mathcal{O}(2^N)$ in memory and $\mathcal{O}(2^{2N})$ per matrix-vector product, while NQS-TDVP requires $\mathcal{O}(\alpha N^2)$ parameters and $\mathcal{O}(n_{\mathrm{samples}} \cdot \alpha N^2)$ per step (p-tVMC multiplies this by $n_{\mathrm{iter}}$). From four data points, the number of required timesteps clearly grows faster than linearly in $N$, but the functional form cannot be pinned down; if $\alpha$ stays moderate, the crossover should occur around $N \approx 15$--20. A head-to-head benchmark against restricted state-space methods~\cite{hogben2011spinach} on identical molecules would establish where, if anywhere, NQS offer a genuine advantage over the best classical alternatives.

Liquid-state NMR spectra of medium-sized organic molecules involve $>$20 coupled protons; proteins can have $>$100. Our 2--5 spin benchmarks do not reach those scales, but two observations are relevant.

The correlation-function route to the spectrum (time-propagate, record $\langle \hat{I}_+(t) \rangle$, Fourier-transform) works straightforwardly with NQS dynamics. The MSE values in Table~\ref{tab:scaling} ($10^{-4}$--$10^{-3}$) correspond to spectral distortions well below typical experimental linewidths (${\sim}\,\SI{1}{\hertz}$), so the accuracy would be sufficient if it persists at larger $N$.

The bottleneck is cost. The number of integration steps grows steeply with $N$, and much of it is spent on the chemical-shift term rather than on the couplings. Removing that term by the interaction-frame transformation (Sec.~\ref{sec:interaction}) recovers a useful fraction of the cost (about eightfold for the 3-spin molecule) and improves the seed-to-seed reproducibility of the larger cases, without changing the spectrum. This addresses only the chemical-shift contribution, however; the cost of representing and integrating strongly entangled states, which we expect to dominate for larger and more densely coupled molecules, is unaffected.

\subsection{Open Questions}

The key open question is how entanglement grows with the size of the largest \emph{connected} coupling cluster. For the molecules studied here, the entanglement within any connected cluster stays at a few bits (Fig.~\ref{fig:entanglement}), but a molecule with 10--20 protons in a single connected network could push the state toward a volume-law regime where exponentially many hidden units are needed. Characterizing that growth would clarify whether RBMs remain the right ansatz or whether transformer-based~\cite{viteritti2023transformer} or autoregressive~\cite{hibatallah2020recurrent} architectures, which have outperformed RBMs in some ground-state problems, are needed for dynamics as well.

Real NMR experiments also include $T_1$ and $T_2$ relaxation, which damp high-frequency oscillations and shorten the effective simulation window. Incorporating Lindblad dynamics~\cite{breuer2002theory} might reduce the number of timesteps needed while making the simulation more physically realistic.

\section{Conclusions}

For 2--5 spin molecular systems, NQS with an RBM ansatz reproduces $^1$H NMR spectra to spectral MSE of $10^{-4}$--$10^{-3}$ at optimal parameters (Table~\ref{tab:scaling}), using both TDVP and p-tVMC propagation. Multi-seed statistics show that initialization sensitivity varies strongly between systems (coefficient of variation from 3\% at $N = 2$ to 32\% at $N = 4$), so reporting single-seed results can be misleading.

At these system sizes, the main finding is that the bottleneck is not network expressivity or memory, but the steep growth in required integration steps with $N$. RBMs with $\alpha \leq 10$ have sufficient representational capacity; what limits accuracy is the accumulation of variational and integration error over thousands of propagation steps. p-tVMC avoids the ill-conditioning issues of QGT inversion and provides a per-step error diagnostic, but at $10$--$20{\times}$ the wall-clock cost of TDVP.

Part of the timestep cost can be removed for free by propagating in the interaction frame of the chemical-shift Hamiltonian, an exact transformation that leaves the spectrum unchanged. In our tests, this cut the steps needed at fixed accuracy roughly eightfold for the 3-spin molecule and by at least an order of magnitude for the 4- and 5-spin systems (Table~\ref{tab:interaction}), and allowed a 14-spin molecule to be propagated by sampling; the interaction-frame results were reproducible across random seeds, in contrast to the rotating-frame ones. It does not change the underlying scaling: the cost of representing strongly entangled states is untouched and will, we expect, set the practical limit for larger and more densely coupled molecules. Natural next steps are to carry the same idea into p-tVMC, applying the offset evolution exactly and projecting only the coupling, and to replace the dense propagator with a Trotter decomposition.

\begin{acknowledgments}
The authors acknowledge computational resources provided by Fraunhofer IAO. This work was supported by the KQCBW25 project, funded by the Ministry of Economic Affairs, Baden-W\"urttemberg. The authors thank the NetKet development team for their open-source contributions to variational Monte Carlo methods. B. X. is supported by the Agency for Science, Technology and Research (A*STAR) International Fellowship.
\end{acknowledgments}

\section*{Data Availability}
The simulation scripts and the cached numerical data underlying all figures and tables are available from the corresponding author upon reasonable request.

\bibliography{references}

\end{document}